\documentclass[aps,pra,twocolumn,letterpaper,showpacs,superscriptaddress,floatfix,10pt,nofootinbib]{revtex4-2}
\usepackage[pdftex]{graphicx}
\graphicspath{ {./images/} }
\usepackage{dcolumn}
\usepackage{bm}
\usepackage{bbm}
\usepackage[usenames, dvipsnames]{color}
\usepackage{comment}
\usepackage[%
  pdfstartview=FitH,%
  breaklinks=true,%
  bookmarks=true,%
  colorlinks=true,%
  anchorcolor=black,%
  citecolor=blue,
  filecolor=black,%
  menucolor=black,%
  urlcolor=blue,%
  linkcolor=blue,%
 ]{hyperref}
\usepackage{layouts}

\usepackage[T1]{fontenc}
\usepackage{exscale}
\usepackage{amsbsy}
\usepackage{subfigure}
\usepackage{textcomp}
\usepackage{physics}
\usepackage{mathtools}
\usepackage{amsfonts}
\usepackage{amssymb}
\usepackage[normalem]{ulem}
\usepackage{amsmath}
\usepackage{mathtools}
\usepackage{tensor}

\usepackage{amsthm}
\theoremstyle{remark}

\usepackage{multirow}

\usepackage[scr=boondoxo, scrscaled=1.05]{mathalfa}

\usepackage{subfigure}
\usepackage{overpic}

\usepackage{array}
\newcolumntype{L}[1]{>{\raggedright\let\newline\\\arraybackslash\hspace{0pt}}m{#1}}
\newcolumntype{C}[1]{>{\centering\let\newline\\\arraybackslash\hspace{0pt}}m{#1}}
\newcolumntype{R}[1]{>{\raggedleft\let\newline\\\arraybackslash\hspace{0pt}}m{#1}}
\usepackage{tabularx}

\def\KeyWord#1{$\backslash$\IfColor{$\!\!$\textRed{#1}\textBlack}{#1}$\!\!$}

\usepackage{wasysym}

\newcommand{\figref}[2]{\hyperref[#1]{\autoref*{#1}(#2)}}
\newcommand{\aref}[1]{\hyperref[#1]{App.~\ref*{#1}}}

\def\tr#1{\mathrm{Tr}\left[#1\right]}

\makeatletter
\let\oldtheequation\theequation
\renewcommand\tagform@[1]{\maketag@@@{\ignorespaces#1\unskip\@@italiccorr}}
\renewcommand\theequation{(\oldtheequation)}
\makeatother

\begin{document}
\title{Probing Hilbert space fragmentation using controlled dephasing}

\author{Dominik Vuina}
\email{dominikv@bu.edu}
\affiliation{Department of Physics, Boston University, Boston, Massachusetts 02215, USA}
\author{Robin Sch\"afer}
\affiliation{Department of Physics, Boston University, Boston, Massachusetts 02215, USA}
\author{David M. Long}
\affiliation{Department of Physics, Stanford University, Stanford, California 94305, USA}
\author{Anushya Chandran}
\affiliation{Department of Physics, Boston University, Boston, Massachusetts 02215, USA}
\affiliation{Max-Planck-Institut für Physik komplexer Systeme, 01187 Dresden, Germany}

\begin{abstract}
    Dynamical constraints in many-body quantum systems can lead to Hilbert space fragmentation, wherein the system’s evolution is restricted to small subspaces of Hilbert space called Krylov sectors.
    However, unitary dynamics within individual sectors may also be slow or non-ergodic, which limits experiments' ability to measure the properties of the entire sector.
    We show that additional controlled dephasing reliably mixes the system within a single Krylov sector, and that simple observables can differentiate these sectors.
    For example, in the strongly interacting XXZ chain with dephasing, the spin imbalance between even and odd sublattices distinguishes sectors. 
    For appropriate choices of initial states, the imbalance begins positive, decays to a negative minimum value at intermediate times, and eventually returns to zero.
    The minimum reflects the average imbalance of the Krylov sector associated to the initial state.
    We compute the size of the minimum analytically in the limit of strong interactions, and validate our results with simulations at experimentally relevant interaction strengths.
\end{abstract}

\maketitle

\section{Introduction}
Thermalization in isolated quantum systems and the processes that prevent it are at the forefront of the study of non-equilibrium physics. While broad classes of many-body systems thermalize rapidly~\citep{PhysRevE.50.888_ETH, d2016quantum_ETH, deutsch2018eigenstate_ETH}, several mechanisms can arrest or dramatically slow the approach to thermal equilibrium. Our work focuses on one such mechanism---Hilbert space fragmentation (HSF)~\citep{garrahan2018aspects_HSF, de2019dynamics, khemani2020localization_HSF, PhysRevX.10.011047_HSF, doi_HSF,PhysRevB.102.195150_HSF_ergodic_Kry_res, feldmeier2020anomalous_HSF, morningstar2020kinetically_HSF, singh2021subdiffusion_HSF, Hahn_2021, PhysRevB.103.134207_HSF, khudorozhkov2022hilbert_HSF, moudgalya2022hilbert_HSF, PhysRevB.106.214426_HSF, 10.21468/SciPostPhys.15.3.093, li2023hilbert_HSF, aditya2024subspace_HSF, yang2025probing_HSF, HSF_exp_2D, scherg2021observing_Exp_1d_hubbard, kohlert2023exploring_1d_hubbard, zhao2025observation_rydberg}---wherein constrained dynamics results in a separation of the Hilbert space into exponentially many (in system size) disconnected blocks, the so-called Krylov sectors. Other mechanisms for the slowdown of thermalization are known, for example: integrability~\citep{Polkovnikov_2011, essler2016quench_int}, many-body scars~\citep{Bernien_2017, moudgalya2022quantum_QMBS, chandran2023quantum_QMBS}, many-body localization~\citep{nandkishore2015many, RevModPhys.91.021001}, and the recently developed many-body caging~\citep{ben2025many, tan2025interferencecagedquantummanybodyscars}. 

Properties of fragmented systems have been extensively theoretically investigated in several model systems~\citep{garrahan2018aspects_HSF, de2019dynamics, PhysRevX.9.021003, khemani2020localization_HSF, PhysRevX.10.011047_HSF, doi_HSF, PhysRevLett.124.207602_HSF, zadnik2021folded, Hahn_2021, PhysRevB.106.214426_HSF, khudorozhkov2022hilbert_HSF, moudgalya2022hilbert_HSF, 10.21468/SciPostPhys.15.3.093, yang2025probing_HSF}.
Fragmented systems do not obey the eigenstate thermalization hypothesis~\citep{PhysRevE.50.888_ETH, d2016quantum_ETH, deutsch2018eigenstate_ETH}; that is, their individual eigenstates do not reproduce thermal expectation values~\citep{garrahan2018aspects_HSF, de2019dynamics, PhysRevX.10.011047_HSF, PhysRevLett.124.207602_HSF,PhysRevB.103.134207_HSF, PhysRevResearch.3.023176, PhysRevB.103.134207_HSF, doi_HSF, Hahn_2021, khudorozhkov2022hilbert_HSF, li2023hilbert_HSF, 10.21468/SciPostPhys.15.3.093, aditya2024subspace_HSF, yang2025probing_HSF}.
Consequently, in dynamical experiments, local observables need not thermalize to their expected Gibbs value~\citep{garrahan2018aspects_HSF, PhysRevX.10.011047_HSF, morningstar2020kinetically_HSF, Hahn_2021, li2023hilbert_HSF, 10.21468/SciPostPhys.15.3.093, aditya2024subspace_HSF, yang2025probing_HSF}.
Transport is also slow in some fragmented systems, being sub-diffusive~\citep{de2019dynamics, PhysRevResearch.2.033124, Feldmeier_2020} instead of diffusive.

Experimentally, fragmented models have been realized in cold atomic systems~\citep{scherg2021observing_Exp_1d_hubbard, kohlert2023exploring_1d_hubbard, HSF_exp_2D} and Rydberg arrays~\citep{zhao2025observation_rydberg}, and HSF has been probed in quench dynamics through macroscopic observables, such as sublattice occupation imbalance (or staggered magnetization), starting from simple initial states in different Krylov sectors. Although the slower relaxation of observables in smaller Krylov sectors is evident in these studies, they are unable to directly measure the properties of entire Krylov sectors. For example, Ref.~\citep{kohlert2023exploring_1d_hubbard} pointed out the difficulty of accessing the steady state imbalance value of the Krylov sector, either due to slow thermalization or non-ergodic dynamics within the sector. 

In order to counter the experimental challenges in probing fragmentation using the dynamics of observables, we propose exposing fragmented systems to controlled dephasing. Many fragmented models in the literature are fragmented in the computational basis, so dephasing does not cause transitions between Krylov sectors.
We show how the average values of observables in individual Krylov sectors can be measured in quench dynamics on short timescales with realistic experimental setups subject to dephasing.
By identifying observables with distinct average values in different Krylov sectors, directly distinguishing Krylov sectors becomes possible.
Further, any potential issues of non-ergodicity within a sector are bypassed, as all initial states approach a uniform mixture of states in their Krylov sector.

Concretely, we demonstrate our proposal by computing imbalance dynamics in the zero-magnetization sector of the spin-1/2 XXZ model with strong interactions and subject to global dephasing in the $z$-basis. The imbalance $I$ is given by the staggered magnetization 
\begin{equation}\label{eq:imbalance}
    I(t) = \frac{2}{L}\sum_{j=1}^L (-1)^j \langle S^{z}_j(t) \rangle,
\end{equation}
with $L$ being the length of the chain, $S^{z}_j$ being the \(z\)-projection of the spin, and the front factor $2/L$ normalizing the imbalance to give $\pm 1$ in the antiferromagnetic N\'eel state. In the limit of infinite nearest neighbor $zz$ interactions, the model is fragmented (and it is called the folded XXZ model in this limit~\citep{PhysRevLett.124.207602_HSF, zadnik2021folded, SciPostPhys.10.5.099}). The Hilbert space splits into distinct Krylov sectors, each with an analytically predictable average imbalance that can be non-zero. 
The XXZ model at strong interactions is proximate for the fragmented model. 
Both strong interactions~\citep{RevModPhys.80.885, scherg2021observing_Exp_1d_hubbard, kohlert2023exploring_1d_hubbard} and controlled dephasing~\citep{PhysRevA.82.063605, PhysRevA.90.023618} can be achieved in ultracold atom setups, with the latter utilizing atom position measurements. 

\begin{figure}
    \centering
    \includegraphics[width=\linewidth]{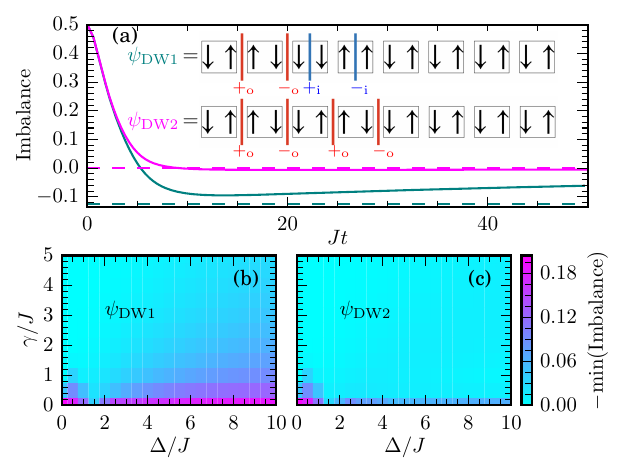}
    \caption{(a) Imbalance decay in the dephased XXZ chain, starting from the two initial states $\psi_{\mathrm{DW1}}$ and $\psi_{\mathrm{DW2}}$, with different domain wall patterns atop the N\'eel state. The average imbalance of the corresponding Krylov sectors in the infinite interaction $(\Delta \to \infty)$ limit, $-1/8$ and $0$, are shown by the teal and pink dashed lines, respectively. The imbalance overshoot, defined as the minimum of the imbalance curve is close to the dashed lines. (b), (c) The absolute value of the imbalance minimum as a function of the model parameters $(\Delta/J, \gamma/J)$ for initial states $\psi_{\mathrm{DW1}}$ and $\psi_{\mathrm{DW2}}$. In our model $\Delta$, $\gamma$ and $J$ are interaction, dephasing and hopping strengths.
    When $\Delta \gg \gamma \gg J$, the size of the overshoot approaches the average imbalance of Krylov sectors of $\psi_{\mathrm{DW1}}$ and $\psi_{\mathrm{DW2}}$, respectively. \emph{Parameters in (a)}: $\gamma=J$, $\Delta=10J$ and $L=16$ obtained by exact diagonalization~\citep{schafer2025danceq,DanceQ_code} of the Lindbladian~\autoref{eq:Lindblad_equation}.}
    \label{fig:phase_diag}
\end{figure}

\figref{fig:phase_diag}{a} shows the quench dynamics of the imbalance with strong but finite interactions beginning from two states belonging to Krylov sectors with distinct average imbalances, $\psi_{\mathrm{DW1}}$ and $\psi_{\mathrm{DW2}}$. The sectors of these states are each characterized by their pattern of domain walls (ferromagnetic bonds) atop the N\'eel state. 
In both states, the imbalance eventually approaches the steady-state value of zero on a timescale controlled by the interaction strength.
Strikingly, as \(\psi_{\mathrm{DW1}}\) has a positive imbalance, but belongs to a sector with a negative average imbalance, $I(t)$ shoots past zero in \figref{fig:phase_diag}{a} on a timescale independent of interaction strength.
The resulting strictly negative minimum in the imbalance as a function of time is close to the Krylov sector average for strong interactions, making the minimum value a distinct and easily measurable signature of the Krylov sector. As the Krylov sector that $\psi_{\mathrm{DW2}}$ belongs to is characterized by a zero average imbalance, the imbalance starting from $\psi_{\mathrm{DW2}}$ shows no such large minimum. 

Panels in \autoref{fig:phase_diag} show the dependence of the imbalance minimum value on the parameters of the model: the interaction strength $\Delta$, dephasing rate $\gamma$, and hopping rate $J$. In the $\Delta \gg \gamma \gg J$ limit, intra-Krylov-sector mixing is much more rapid than inter-Krylov-sector mixing, so that the minimum value is close to the average imbalance of the associated Krylov sector at $\Delta \to \infty$. When $\gamma \approx 0$ and $\Delta \gg J$, the imbalance coherently oscillates around the same average value. This results in a larger (in magnitude) imbalance minimum near the $\gamma=0$ line as compared to the $\Delta \gg \gamma \gg J$ limit. The panels show that there is an extended parameter regime where the average imbalance of the associated Krylov sector can be extracted from the imbalance minimum. The limit of weak interactions $\Delta \ll J$ is not of interest because it is not fragmented.

We organize this article as follows. First, we describe the XXZ model with global dephasing in \autoref{sec:model}. In \autoref{sec:inf_int_limit}, we explain the Krylov sector structure in the \(\Delta \to \infty\) XXZ model, as well as the effective dynamics in the strongly interacting limit of the model with and without strong dephasing. \autoref{sec:results} demonstrates how the quench dynamics in the dephased limit can be used to probe Krylov sectors. Finally, we dedicate \autoref{sec:strong_int_strong_deph} to calculating the properties of the Krylov sectors, namely their dimension and average imbalance.

\section{Model}\label{sec:model}

We consider the dynamics of the spin-1/2 XXZ model on $L$ sites with periodic boundary conditions. The Hamiltonian is given by

\begin{equation}\label{eq:Hamiltonian}
    H=\sum_{j=1}^{L}J\left(S^{x}_jS^{x}_{j+1} + S^{y}_jS^{y}_{j+1}\right) + \Delta S^{z}_jS^{z}_{j+1},
\end{equation}
where $S^{\alpha}_j$ are the spin-1/2 operators with $\alpha\in\{x,y,z\}$ on site $j$, $\Delta, J > 0$, and periodic boundary conditions identify the site $L+1$ with the site $1$. The symmetries of the model are translational invariance, global spin-rotation about the $z$-axis, and a global spin-flip about the \(x\), \(y\), and \(z\) axes. The global $U(1)$ symmetry splits the XXZ model into distinct magnetization sectors with conserved total spin \(z\)-component $S^{z}_{\mathrm{tot}}=\sum_j S^{z}_j$. This conservation law plays an important role in the formation of Krylov sectors and our results. The global spin-flip and translational invariance can both be broken without influencing our conclusions. In this manuscript, we focus on the $S^{z}_{\mathrm{tot}}=0$ magnetization sector.

Our primary model of interest is the XXZ model in the presence of global dephasing. The dynamics of the density matrix of the system $\rho$ are captured by the Lindblad equation~\citep{breuer2002theory, manzano2020short}
\begin{align}
    \frac{\mathrm{d}\rho}{\mathrm{d}t} = \mathcal{L}[\rho] = &-i[H, \rho] + \sum_{j=1}^L l_j \rho l_j^{\dagger} - \frac{1}{2}\left\{ l_j^{\dagger} l_j, \rho \right\}, \label{eq:Lindblad_equation}\\
    & \mathrm{with} \quad l_j = \sqrt{\gamma}S^{z}_j,\label{eq:disspative_ops}
\end{align}
where $\gamma$ is the strength of dephasing in the $z$-basis, and we set $\hbar=1$. Global dephasing decoheres the system in the $z$-product basis. 

HSF is consistent with dephasing in the $\Delta\to\infty$ fragmented limit of the XXZ model. A density matrix initialized in some Krylov sector remains in the same sector, and dephasing only exponentially damps out its off-diagonal elements in the $z$-basis. 

In the following sections, we consider the quench dynamics of \autoref{eq:Lindblad_equation} starting from two initial states $\psi_{\mathrm{DW1}}$ and $\psi_{\mathrm{DW2}}$. These are $z$-basis spin configurations depicted in \figref{fig:phase_diag}{a} at $L=16$. We take the thermodynamic limit by repeating the $L=16$ pattern of these configurations.

\section{Large interaction strength}\label{sec:inf_int_limit}

In the limit of infinite interaction strength $\Delta \to \infty$, the XXZ model is fragmented into exponentially many (in $L$) Krylov sectors. In the zero magnetization sector $S^{z}_{\mathrm{tot}}=0$, each sector is labelled by a fixed pattern of domain walls atop one of the N\'eel states. When $\Delta/J \to \infty$, the two N\'eel states span the groundspace of the model and the excitations are domain walls---ferromagnetic bonds between neighboring spins. We pick the N\'eel state with imbalance $I=1$ according to \autoref{eq:imbalance} as a reference. The domain walls are nucleated by spin exchange on top of a Néel state. Creating or annihilating domain walls incurs an energy penalty $\Delta$, so sectors with different numbers of domain walls are dynamically disconnected as \(\Delta \to \infty\).

\begin{figure}
    \centering
    \includegraphics[width=\linewidth]{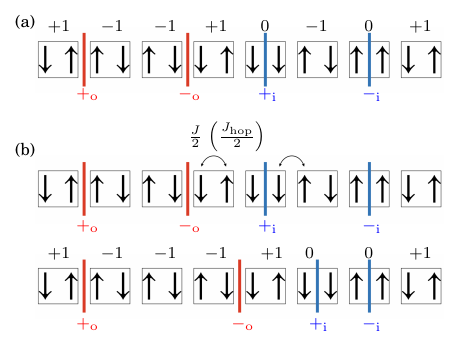}
    \caption{(a) A configuration of the spin chain in the $z$-basis with $L=16$ sites and two pairs of domain walls. The subscript refers to the domain walls inside $(\mathrm{i})$ or outside $(\mathrm{o})$ the unit cell, and $\pm 1$ refers to the imbalance of the unit cell to the left of the domain wall. The imbalance of each unit cell is written above it. (b) The top configuration illustrates a few allowed spin exchanges in the $\Delta \to \infty$ limit, occurring at the rate $J/2 \, (J_{\mathrm{hop}}/2)$ in the unitary (strongly dephased, defined in \autoref{eq:noninteracting_rate}) limits. In the lower configuration, the $-_{\mathrm{o}}$ and $+_{\mathrm{i}}$ domain walls have hopped to the right. The hop of the $-_{\mathrm{o}}$ domain wall increases the size of the $-1$ imbalance domain between the $\pm_{\mathrm{o}}$ walls. Similarly, when $+_{\mathrm{i}}$ hops the length of the $-1$ imbalance region between the $\pm_{\mathrm{i}}$ walls decreases.}
    \label{fig:dw_label_hops}
\end{figure}

To each $z$-basis product state, we assign a \emph{domain wall string}, as follows. We pair sites into two-site unit cells. A domain wall is labeled by: whether it is inside or outside a unit cell ($\mathrm{i}$ or $\mathrm{o}$); the imbalance of a domain-wall-free unit cell to its left (\(\pm\)); and its position on the \(r\)th bond of the chain. The allowed labels are thus $\pm_{\mathrm{i}_{(r)}}$ and $\pm_{\mathrm{o}_{(r)}}$. Any configuration with $S^{z}_{\mathrm{tot}}=0$ (except the two N\'eel states) can be uniquely mapped into a domain wall string. The empty string is an exception as it describes both N\'eel states.  For example, `$+_{\mathrm{o}_{(2)}},-_{\mathrm{o}_{(6)}}, +_{\mathrm{i}_{(9)}}, -_{\mathrm{i}_{(13)}}$' is the string for the spin configuration in \figref{fig:dw_label_hops}{a}. Domain walls act as hard-core particles---that is, they cannot occupy the same bonds (\figref{fig:dw_label_hops}{b}). 

All spin configurations in the same Krylov sector share the same \emph{reference string}, defined as a string without position indices and up to cyclic permutations (to account for periodic boundary conditions). 
For example, all configurations in \autoref{fig:dw_label_hops} are in the sector labelled by the reference string `$+_{\mathrm{o}},-_{\mathrm{o}}, +_{\mathrm{i}}, -_{\mathrm{i}}$', or equivalently its cyclic permutation `$-_{\mathrm{o}}, +_{\mathrm{i}}, -_{\mathrm{i}}, +_{\mathrm{o}}$'. 

Valid domain wall strings that correspond to $S^{z}_{\mathrm{tot}}=0$ configurations have the following two properties. Firstly, adjacent domain walls have alternating imbalance signs. Secondly, with an appropriate assignment of brackets to the domain walls, only those strings that are Dyck words~\citep{shapiro1976catalan}---with an equal number of brackets of the same type (for example angled and round), are valid $S^{z}_{\mathrm{tot}}=0$ configurations. This is a consequence of domain walls being nucleated in pairs. We assign $+_{\mathrm{i}}$ and $+_{\mathrm{o}}$ domain walls to open round $($ and angled $[$ parentheses, respectively. Likewise, closed round $)$ and angled $]$ parentheses represent $-_{\mathrm{i}}$ and $-_{\mathrm{o}}$ walls. The rule for alternating the imbalance sign of domain walls implies that all open parentheses must be adjacent to closed parentheses.
The reference strings of configurations in \autoref{fig:dw_label_hops} correspond to Dyck words `()[]'. The reference string `[)[]()[)' is not a Dyck word since the same kind of parentheses are not balanced; a small change to the string `(][]()[)' makes it a valid $S_{\mathrm{tot}}^z=0$ string (for periodic boundary conditions).

In the following sections, we describe the dynamics of domain walls restricted to the Krylov sectors. The domain walls hop coherently at $\gamma=0$. For $\gamma > 0$, the late-time dynamics for domain walls are governed by an incoherent hopping model. 

\subsection{Closed system}\label{sec:inf_int_limit_unitary}

When \(\gamma=0\), there is an effective Hamiltonian, obtained by perturbation theory in \(J/\Delta\), which captures the dynamics of domain walls. At leading (constant in \(J/\Delta\)) order, it is given by~\citep{zadnik2021folded, SciPostPhys.10.5.099} 
\begin{equation}\label{eq:hopping_unitary}
    H_{\Delta \to \infty} = \frac{J}{2}\sum_{j}\left(1+4S^{z}_{j-1}S^{z}_{j+2}\right)\left(S^{x}_jS^{x}_{j+1} + S^{y}_jS^{y}_{j+1} \right).
\end{equation}

\autoref{eq:hopping_unitary} allows domain walls to hop left or right by two sites (one unit cell) with amplitude $J/2$. 
At large but finite interaction strength $\Delta \gg J$, \autoref{eq:hopping_unitary} describes the dynamics of the XXZ model on time scales much less than $\Delta/J^2$, which is the time scale on which the next term in perturbation theory becomes important.

\subsection{Open System}\label{sec:inf_int_dephas}

When dephasing is large, $\gamma \gg J$, the dynamics of \autoref{eq:Lindblad_equation} on timescales $t \gg 1/\gamma$ are captured by a model with only incoherent transitions between \(z\)-basis states for any $\Delta$. As $\gamma \to \infty$, every operator $|\phi_i\rangle\langle \phi_i|$, made up of the $z$-product states $|\phi_i\rangle$, is a steady state. At finite $\gamma \gg J$, second order perturbation theory provides transition rates between diagonal ensembles in the $z$-basis~\citep{PhysRevB.93.094205}, which can be viewed as transitions between the $z$-basis configurations of the chain.

\begin{figure}
    \centering
    \includegraphics[width=\linewidth]{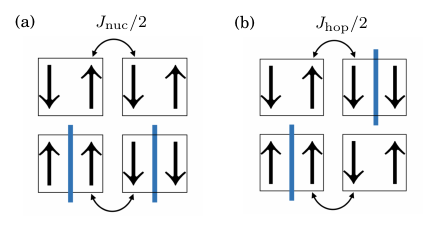}
    \caption{Illustration of the hopping terms in the effective model given by \autoref{eq:effective_Hamiltonian}, with strengths $2J_{\mathrm{nuc}}$ and $2J_{\mathrm{hop}}$, respectively. In (a), spin exchange creates (top) or annihilates (bottom) a pair of $\mathrm{i}$ domain walls. In (b), the same process of spin exchange hops an $\mathrm{i}$ domain wall. }
    \label{fig:ji_jn_hops}
\end{figure}

The spin exchanges that nucleate or annihilate domain walls  (\figref{fig:ji_jn_hops}{a}) occur at the rate $J_{\mathrm{nuc}}/2$, which depends on the interaction strength,
\begin{equation}\label{eq:interacting_rate}
    J_{\mathrm{nuc}}=J^2\gamma/(\gamma^2 + \Delta^2 ).
\end{equation}
For large interaction strengths $\Delta \gg \gamma$, these transitions occur at a small rate $J_{\mathrm{nuc}} \sim J^2 \gamma/\Delta^2$, reflecting proximity to the fragmented model where the number of domain walls is conserved. 
Meanwhile, the spin exchanges that hop the domain walls (\figref{fig:ji_jn_hops}{b}) occur at a $\Delta$-independent rate $J_{\mathrm{hop}}/2$, given by
\begin{equation}\label{eq:noninteracting_rate}
    J_{\mathrm{hop}} = J^2/\gamma. 
\end{equation}
Domain walls hop often and are nucleated/annihilated rarely as $J_{\mathrm{nuc}} \ll J_{\mathrm{hop}}$ in the limit of strong interactions.

The perturbative treatment of \autoref{eq:Lindblad_equation} in the large dephasing limit $J/\gamma \to 0$ generates the equation of motion for the vectorized diagonal part of the density matrix~\citep{horn1994topics, machnes2014surprisinginteractionsmarkoviannoise} $|\rho_D) = \sum_{i=1}^{D_{0}} p_i(t)|\phi_i)$ given by
\begin{equation}
    \frac{\mathrm{d}|\rho_D)}{\mathrm{d}t}=-H_{\mathrm{eff}}|\rho_D),\label{eq:eff_evolution}
\end{equation}
where $|\phi_i) = |\phi_i\rangle \otimes |\phi_i\rangle$ in the notation for the diagonal density matrix in a doubled Hilbert space, and $D_{0}$ is the dimension of the zero magnetization sector.
The effective Hamiltonian is given by~\citep{PhysRevB.93.094205} 
\begin{multline}\label{eq:effective_Hamiltonian}
        H_{\mathrm{eff}}=\frac{1}{4}\sum_{\alpha,\beta=\pm 1}\sum_{j=1}^L \frac{J_{\alpha,\beta}}{4}\left(1+2\alpha \tilde{S}^{z}_{j-1}\right)\left(1-4\vec{\tilde{S}}_{j}\cdot\vec{\tilde{S}}_{j+1}\right) \\
    \times\left(1+2 \beta \tilde{S}^{z}_{j+2}\right),
\end{multline}
where $\vec{\tilde{S}}_{j}\cdot\vec{\tilde{S}}_{j+1} = \tilde{S}^z_j \tilde{S}^z_{j+1} + \left( \tilde{S}^{+}_j \tilde{S}^{-}_{j+1} + \tilde{S}^{-}_j \tilde{S}^{+}_{j+1} \right)/2$, $\tilde{S}_j^{\pm} = S_j^{\pm} \otimes S_j^{\pm}$ and $\tilde{S}_j^z= S_j^z \otimes 1$  . The front factor is $J_{\alpha,\beta}=4J^2\gamma/(4\gamma^2+(\alpha\Delta-\beta\Delta)^2)$, and the boundary conditions are the same as in~\autoref{eq:Hamiltonian}. This model generates transitions between the diagonal density matrix elements as depicted in \autoref{fig:ji_jn_hops}, with $J_{\mathrm{nuc}} = J_{+,-}=J_{-,+}=J^2\gamma/\left(\gamma^2 + \Delta^2\right)$ and $J_{\mathrm{hop}} = J_{+,+}=J_{-,-}=J^2/\gamma$. While \autoref{eq:hopping_unitary} describes coherent hopping of domain walls, \autoref{eq:effective_Hamiltonian} describes incoherent hopping. 

The evolution of the diagonal part of the density matrix, as described in \autoref{eq:eff_evolution}, can be interpreted as imaginary-time evolution under an effective Hamiltonian, applied to the initial (diagonal) density matrix. This effective evolution maintains the positivity of the density matrix and preserves the trace. In the long-time limit $t \to \infty$, the vectorized density matrix approaches the ground state $|\Psi_{\rm eff})$ of $H_{\mathrm{eff}}$, which corresponds to the steady state of the system. This state is a maximally mixed state with eigenvalue zero, satisfying $(1 - 4\vec{\tilde{S}}_j \cdot \vec{\tilde{S}}_{j+1}) |\Psi_{\rm eff}) = 0$ for all $j$.
This prediction of the steady state from the effective model is in agreement with the steady state of the full Lindblad equation \autoref{eq:Lindblad_equation}, which is the identity matrix.

In the limit $\Delta \to \infty$, the nucleation rate vanishes $J_{\mathrm{nuc}} = 0$, and both models (full Lindbladian and the effective Hamiltonian) exhibit a large degeneracy of steady states---one for each Krylov sector. Each of these steady states is a uniform mixture over the basis states of its respective Krylov sector.

\section{Probing Hilbert Space Fragmentation}\label{sec:results}

We demonstrate how controlled dephasing gives access to observable Krylov sector properties in the dynamics of the strongly interacting XXZ chain. The dynamics are studied starting from initial states $\psi_{\mathrm{DW1}}$ and $\psi_{\mathrm{DW2}}$ with distinct domain wall reference strings, `$+_{\mathrm{o}},-_{\mathrm{o}}, +_{\mathrm{i}}, -_{\mathrm{i}}$' and `$+_{\mathrm{o}},-_{\mathrm{o}}, +_{\mathrm{o}}, -_{\mathrm{o}}$', in distinct Krylov sectors. Quench dynamics of the $\psi_{\mathrm{DW1}}$ and $\psi_{\mathrm{DW2}}$ initial states indicate an asymptotic separation of timescales between the time required to mix within Krylov sectors and the time required to relax towards the maximally mixed state. We observe this separation both in the von Neumann entropy, $S(t)=-\tr{\rho(t) \log \rho(t)}$, and the imbalance expectation value, \autoref{eq:imbalance}. The imbalance overshoot value, defined as the minimum value of imbalance as a function of time, approaches the average imbalance of the initial state's Krylov sector in the $\Delta \to \infty$ limit. Thus, the imbalance overshoot for $\Delta \gg J$ is a signature of HSF as $\Delta \to \infty$, and further reveals a property of the entire Krylov sector.  

By contrast, in the strongly interacting limit $\Delta \gg J$ and $\gamma=0$, the dynamics do not simply reflect properties of Krylov sectors. Both $\psi_{\mathrm{DW1}}$ and $\psi_{\mathrm{DW2}}$ have energy densities that correspond to the infinite temperature state restricted to their respective Krylov sector (with respect to \autoref{eq:hopping_unitary}). Starting from these initial states, we find that the imbalance exhibits substantial oscillations, obscuring the measurement of the average imbalance.

\subsection{Dynamics with Strong Dephasing}\label{sec:results_deph}

We focus on timescales $t \gg 1/\gamma$ in the strong dephasing limit, where the effective model~\autoref{eq:effective_Hamiltonian} governs the dynamics. 

Consider first the infinite interaction strength limit. Since the steady state is the uniform mixture of all $z$-basis configurations in the initial state's Krylov sector, the von Neumann entropy plateaus at $S(t\to\infty)=\log(D_K)$, where $D_K$ is the dimension of the sector. The steady state imbalance $I_{\mathrm{ss}}$ is given by the imbalance uniformly averaged over the sector. In \autoref{sec:imbalance}, we calculate the dimensions and average imbalances of all Krylov sectors in the $S_{\mathrm{tot}}^z=0$ sector. At system size $L=16$, the two states $\psi_{\mathrm{DW1}}$ and $\psi_{\mathrm{DW2}}$ have the steady state imbalances $I_{\mathrm{ss}}=-1/8$ and $I_{\mathrm{ss}}=0$, respectively (\autoref{sec:imbalance}). Note that fixing the reference string of $\psi_{\mathrm{DW1}}$ and increasing $L$ yields $I_{\mathrm{ss}} = \order{L^{-1}}$ scaling as $L\to\infty$. In the thermodynamic limit, a finite steady state imbalance requires a finite density of domain walls in the initial state. This can be achieved, for instance, by repeating the $L=16$ domain wall patterns of $\psi_{\mathrm{DW1}}$ and $\psi_{\mathrm{DW2}}$.

At large but finite interaction strengths $\Delta \gg \gamma$, there is a separation of timescales in the time evolution of the model in \autoref{eq:effective_Hamiltonian}. On timescales $T_{\mathrm{hop}}= \mathcal{O}(J_{\mathrm{hop}}^{-1})$, initial states mix primarily within their Krylov sectors. On timescales $T_{\mathrm{nuc}}=\mathcal{O}(J_{\mathrm{nuc}}^{-1})$, when the probability of domain wall nucleation and annihilation is finite, the distinct Krylov sectors mix. This leads to all initial states evolving to the maximally mixed steady state in the appropriate $S^z_{\mathrm{tot}}$ sector.

\begin{figure}
    \centering
    \includegraphics[width=\linewidth]{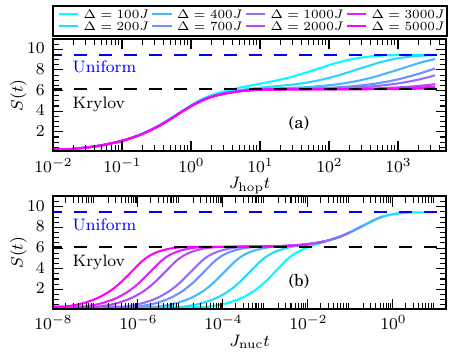}
    \caption{Time dependent von Neumann entropy in the dephased XXZ model, extracted by exactly diagonalizing the effective model~\autoref{eq:effective_Hamiltonian} at different interaction strengths $\Delta/J$. The initial state is $\psi_{\mathrm{DW1}}$. (a) The entropy plateaus at intermediate times and reaches the value $S(t)=\log(D_K)$ (labelled as `Krylov' in the figure, $D_K=448$ at $L=16$), consistent with mixing within the Krylov sector of $\psi_{\mathrm{DW1}}$. At long times, the curves saturate to the uniform value of the zero-magnetization sector (labelled as `Uniform' in the figure), that is $\log D_{0}$. The scaling collapse in (b) shows that the timescale on which the fully mixed state is reached scales with $J_{\mathrm{nuc}}^{-1}$. \emph{Parameters}: $\gamma=3J$ and $L=16$.}
    \label{fig:entropy}
\end{figure}

\autoref{fig:entropy} shows evidence for the separation of timescales. On $t\approx T_{\mathrm{hop}}$ timescales, the von Neumann entropy exhibits a plateau, associated with mixing within the Krylov sector of the initial state $\psi_{\mathrm{DW1}}$. The value of the plateau is given by $S=\log(D_K)$, where $D_K=448$ is the dimension of the Krylov sector that $\psi_{\mathrm{DW1}}$ belongs to (\autoref{sec:dimension}). On timescales $t \approx T_{\mathrm{nuc}}$, the entropy climbs to a second plateau at $\log D_{0}$. Scaling time by \(J_{\mathrm{hop}}\) or \(J_{\mathrm{nuc}}\) produces scaling collapse at short and long times, respectively, confirming the separation of timescales. 
The separation of timescales enables the extraction of properties of the Krylov sectors in the system with dephasing on timescales $T_{\mathrm{hop}} \ll t \ll T_{\mathrm{nuc}}$. For von Neumann entropy, this is the plateau value $S=\log{D_K}$ in \autoref{fig:entropy}. For imbalance, this is the overshoot size, which reflects the average Krylov sector imbalance $I_{\mathrm{ss}}$, as we now show. 

\begin{figure}
    \centering
    \includegraphics[width=\linewidth]{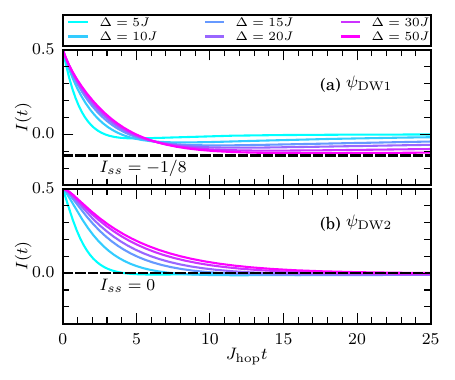}
     \caption{Imbalance in the open system, extracted by exactly diagonalizing the effective model~\autoref{eq:effective_Hamiltonian} at different interaction strengths $\Delta/J$ starting from (a) $\psi_{\mathrm{DW1}}$ and (b) $\psi_{\mathrm{DW2}}$. For increasing interaction strength and $t \approx T_{\mathrm{hop}}$, the imbalance curve approaches the average Krylov sector imbalance $I_{\mathrm{ss}}=-1/8$ in (a) and $I_{\mathrm{ss}}=0$ in (b). Note that at longer times, the curves in (a) approach $I=0$ (see \figref{fig:phase_diag}{a}). \emph{Parameters}: $\gamma=3J$ and $L=16$.}
    \label{fig:origin}
\end{figure}

At large interaction strengths and intermediate timescales, $T_{\mathrm{hop}} \ll t \ll T_{\mathrm{nuc}}$, the imbalance overshoot approaches the values $I_{\mathrm{ss}}=-1/8$ and $I_{\mathrm{ss}}=0$ for states $\psi_{\mathrm{DW1}}$ and $\psi_{\mathrm{DW2}}$, as shown in \autoref{fig:origin}. 

The overshoot timescale---that is, the time at which the imbalance first crosses zero---is set by $T_{\mathrm{hop}}$, which is asymptotically independent of the interaction strength. This can be seen for the $\Delta \gg \gamma$ curves in \figref{fig:origin}{a}, or by noting that the \(\Delta \to \infty\) model exhibits an overshoot. 
For different initial states, we expect the overshoot timescale 
to depend on the domain wall density as $T_{\mathrm{hop}} \rho_{\mathrm{DW}}^{-2}$, where \(\rho_{\mathrm{DW}}\) is the density of domain walls. The dependence on $\rho_{\mathrm{DW}}$ follows from the time it takes the domain walls to diffuse across a length set by the inter-domain spacing. Dimensional analysis then predicts that the time for the overshoot scales as \(D^{-1} \rho_{\mathrm{DW}}^{-2}\), where \(D \propto 1/T_{\mathrm{hop}}\) is the diffusion constant.

The size of the overshoot depends on the imbalance $I_{\mathrm{ss}}$ of the initial state's Krylov sector, which can be calculated from the sector's reference string and the number of lattice sites $L$ (\autoref{sec:imbalance}). 

\begin{figure}
    \centering
    \includegraphics[width=\linewidth]{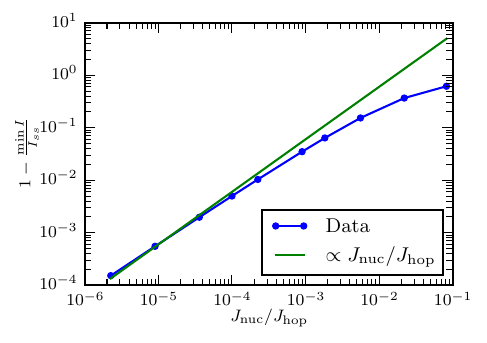}
    \caption{Scaling of the dimensionless imbalance minimum in state $\psi_{\mathrm{DW1}}$, $1-\min I/I_{\mathrm{ss}}$ as a function of $J_{\mathrm{nuc}}/J_{\mathrm{hop}}$, in the $J_{\mathrm{nuc}} \ll J_{\mathrm{hop}}$ limit. The linear scaling relationship can be used to fit the overshoot data for a given initial state at finite interaction strengths and extract its $I_{\mathrm{ss}}$ value.
    Data is obtained from exact diagonalization on the effective model~\autoref{eq:effective_Hamiltonian} in the $\psi_{\mathrm{DW1}}$ initial state. \emph{Parameters}: $\gamma=3J$, $J=1$ and $L=16$.}
    \label{fig:minimum_scale}
\end{figure}

For the $\psi_{\mathrm{DW1}}$ initial state, the absolute value of the imbalance minimum empirically scales as $I_{\mathrm{ss}}(1 - C J_{\mathrm{nuc}}/J_{\mathrm{hop}})$ in the $J_{\mathrm{hop}} \gg J_{\mathrm{nuc}}$ limit, where $C$ is a constant (which we expect to be independent of $L$ at large $L$).
As the rate of domain wall nucleation and annihilation is $J_{\mathrm{nuc}}$ and the overshoot occurs at $t \approx T_{\mathrm{hop}}$, the number of Krylov sectors the initial state has mixed with by the time the overshoot occurs scales as $\propto J_{\mathrm{nuc}}/J_{\mathrm{hop}}$. Mixing within those sectors reduces the imbalance overshoot size by the factor $C J_{\mathrm{nuc}}/J_{\mathrm{hop}}$, with \autoref{fig:minimum_scale} providing evidence for such scaling. 

Besides being easily accessible in experiments, the imbalance overshoot is a more useful probe of Krylov sector properties than the von Neumann entropy at finite interaction strengths $\Delta$. The overshoot approaches $I_{\mathrm{ss}}$ on smaller interaction strengths ($\Delta = 50J$ in \autoref{fig:origin}) compared to the entropy plateau emerging at $S=\log(D_K)$ (about $\Delta = 400 J$ in \autoref{fig:entropy}).

At finite $\Delta$, the overshoot is also simpler to extract from the imbalance as a function of time compared to the plateau in the von Neumann entropy. The approximate plateau at $\Delta=200J$ in \autoref{fig:entropy} is an inflection point. The overshoot is the minimum of the imbalance curve, which is easier to observe compared to this inflection point. 
Some initial states have a positive initial imbalance and are in a sector with positive $I_{\mathrm{ss}}$ (\autoref{sec:imbalance}). Like entropy, imbalance curves then exhibit a plateau at intermediate times. We focus on initial states with an overshoot, as this feature is more striking and is easier to observe. 

Finally, dynamical processes that connect different Krylov sectors also change the imbalance to a smaller extent than the von Neumann entropy. At finite $\Delta$, the nucleation or annihilation of domain walls connects Krylov sectors. These local moves connect sectors that differ in average imbalance by $\order{1/L}$, in contrast to the $\order{\log(L)/L}$ increase in entropy density, which counts the new accessible states. Moreover, since average imbalance values of connected sectors can be positively or negatively different from the initial sector's \(I_{\mathrm{ss}}\), the imbalance may also approach zero more slowly due to cancellations. 

\subsection{Unitary dynamics}\label{sec:results_unitary}

\begin{figure}
    \centering
    \includegraphics[width=\linewidth]{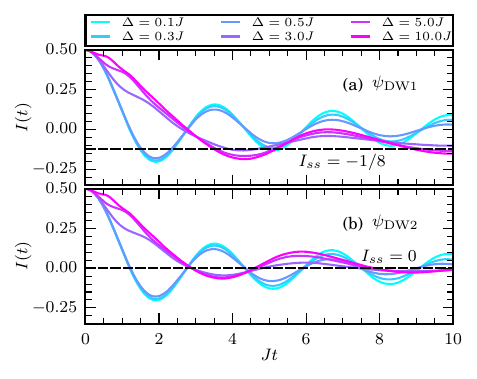}
    \caption{Imbalance dynamics in the unitary XXZ model~\autoref{eq:Hamiltonian} as a function of $\Delta/J$ starting from initial state (a) $\psi_{\mathrm{DW1}}$ and (b) $\psi_{\mathrm{DW2}}$ on system size $L=32$, by twice repeating the $L=16$ pattern of these states (see \figref{fig:phase_diag}{a}). With strong interactions $\Delta \gg J$, the curves exhibit damped oscillations about the long time imbalance values $I_{\mathrm{ss}}=-1/8$ and $I=0$ in the $\psi_{\mathrm{DW1}}$ and $\psi_{\mathrm{DW2}}$ initial states, respectively. The damping rate and oscillation frequency appear to be independent of the interaction strength. The curves are extracted with the TEBD algorithm for time evolution under the Hamiltonian in \autoref{eq:Hamiltonian} with MPS~\citep{SciPostPhysCodeb.4, SciPostPhysCodeb.4r0.3}. A singular value truncation cutoff of $1 \times 10^{-7}$ was used. \emph{Parameters}: $J=1$ and $L=32$.}
    \label{fig:interacting_sys_unitary}
\end{figure}

The imbalance dynamics at $\gamma=0$ and large interaction strengths also show features of HSF. However, at intermediate times, oscillations in the imbalance obscure signatures of HSF in the imbalance dynamics. 
Unlike the case with dephasing, the imbalance at late times can also depend on the initial state (as the XXZ model is integrable), not just on the Krylov sector.  

\autoref{fig:interacting_sys_unitary} shows that when $\Delta \gg J$, the imbalance dynamics starting from the two states $\psi_{\mathrm{DW1}}$ (a) and $\psi_{\mathrm{DW2}}$ (b) is oscillatory approximately around the average values of $I=-1/8$ and $I=0$, respectively.
If the dynamics in these two initial states are ergodic, one could extract \(I_{\mathrm{ss}}\) either by waiting for the oscillations to die down or performing a fit to the oscillating signal.
If the unitary dynamics under \autoref{eq:hopping_unitary} is non-ergodic for any initial state, the addition of dephasing guarantees the approach to $I_{\mathrm{ss}}$. 
More broadly, introducing dephasing provides additional control over the approach to \(I_{\mathrm{ss}}\), beyond the dynamics dictated by \autoref{eq:hopping_unitary}.

Oscillations around $I_{\mathrm{ss}}$ cause the imbalance overshoot to be larger in the unitary case as compared to the strongly interacting dephased XXZ model (as can be seen in \figref{fig:phase_diag}{b} and \figref{fig:phase_diag}{c}). The exact value of the overshoot remains difficult to predict. Despite being Bethe ansatz integrable~\citep{ zadnik2021folded, SciPostPhys.10.5.099}, there is currently no closed-form solution for the quench dynamics of the imbalance in the XXZ model.

\autoref{fig:interacting_sys_unitary} also shows that in the $\Delta \ll J$ regime, without HSF, the imbalance oscillations occur around $I=0$, with an increased frequency (compared to the $\Delta \gg J$ regime). The imbalance curves of the two states $\psi_{\mathrm{DW1}}$ and $\psi_{\mathrm{DW2}}$ in this regime are further indistinguishable. 

The imbalance oscillations in the $\Delta \ll J$ regime result from the coherent hops of spins on the chain. In the $\Delta=0$ non-interacting limit of the XXZ model, the imbalance as a function of time can be computed exactly by a mapping to free fermions (for details, see {\aref{sec:non_int_limit}}). When $t > J^{-1}$, the imbalance exhibits harmonic oscillations. The amplitude of these oscillations is set by the imbalance of the initial state (equal for $\psi_{\mathrm{DW1}}$ and $\psi_{\mathrm{DW2}}$) and it decays as $t^{-1/2}$. 
This phenomenology has a simple interpretation. At $\Delta=0$, the spin excitations (the $L/2$ up spins in the $z$-basis) spread ballistically across the chain. Propagating spins move from the even to the odd sublattice and back periodically, leading to imbalance oscillations at the hopping frequency. We may consider the spin excitations as wavepackets, which spread out due to dispersion, giving a power-law decay of oscillations at \(\Delta=0\). At weak interactions, the spin scattering leads to exponential damping of imbalance oscillations, as reported in Ref.~\citep{barmettler2009relaxation} for the N\'eel initial state. We expect $\psi_{\mathrm{DW1}}$ and $\psi_{\mathrm{DW2}}$ to behave similarly, as the presence of domain walls does not have a strong effect in the small \(\Delta\) limit.

The imbalance oscillation amplitude and frequency differ in the $\Delta \gg J$ and $\Delta=0$ limits. The effective model at large $\Delta$ in \autoref{eq:hopping_unitary} shows that the domain walls are the mobile particles in this limit, rather than the spin up excitations. The comparably smaller density of mobile domain walls may account for the difference in frequency of oscillations in the two limits. Future work could focus on understanding the oscillations in $\Delta \gg J$ limit.

\section{Krylov sector properties}\label{sec:strong_int_strong_deph}
We compute properties of the Krylov sectors of the XXZ model at $\Delta \to \infty$. Using the allowed transitions between the spin configurations in the effective models \autoref{eq:hopping_unitary} and \autoref{eq:effective_Hamiltonian}, we find closed form expressions for the dimension $D_K$ and average imbalance $I_{\mathrm{ss}}$ of all Krylov sectors.

\subsection{Krylov Sector Dimension}\label{sec:dimension}

The dimension of each Krylov sector $D_K$ is (with some exceptions like the N\'eel states) the number of $z$-basis configurations with the same domain wall reference string (see \autoref{sec:inf_int_limit}). 
Any two states with the same non-empty reference string, up to cyclic permutations on periodic boundary conditions, can be connected by hopping domain walls, provided there is at least one \emph{free unit cell} with no domain walls (defined more precisely below). We count the number of states with a given reference string by counting the possible arrangements of free unit cells between the domain walls.

\begin{figure}
    \centering
    \includegraphics[width=\linewidth]{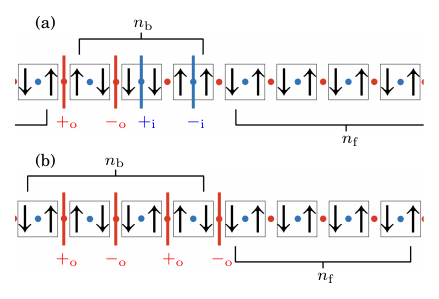}
    \caption{(a) $\psi_{\mathrm{DW1}}$ state with labelled domain walls on the occupied domain wall sites (red and blue dots between spins). There are two $\mathrm{i}$ domain walls, $n_{\mathrm{DW},i}=2$, and one adjacent pair of $\mathrm{o}$ domain walls, $p_{o}=1$, giving three bound unit cells, $n_{\mathrm{b}}=3$, and $n_{\mathrm{f}}=L/2 - n_{\mathrm{b}}=5$ free unit cells. Only the identity operation leaves the $\psi_{\mathrm{DW1}}$ string invariant, so $n_{\mathrm{stab}}=1$. The Krylov sector dimension with $n_{\mathrm{f}}=5$ and $n_{\mathrm{DW}}=4$ is $D_K=448$. 
    (b) $\psi_{\mathrm{DW2}}$ state, with $n_{\mathrm{b}}=4$ cells bound to adjacent pairs of $\mathrm{o}$ walls ($p_{\mathrm{o}}=4$), $n_\mathrm{f}=4$ and $n_{\mathrm{stab}}=2$. Thus, the sector it belongs to has $D_K=140$.}
    \label{fig:dw_sites}
\end{figure}

The domain walls are hard-core particles living on the bonds of the lattice and with short-range repulsion. That is, domain walls of different type can be on neighbouring bonds, whereas the domain walls of the same type are at least one bond apart. This follows from the $\pm_{\mathrm{o}}$ and $\pm_{\mathrm{i}}$ domain walls hopping on even and odd domain wall sublattices, respectively (\autoref{fig:dw_sites}).

We first show how to count the Krylov sector dimension for open boundary conditions. Subsequently, we do the counting with periodic boundary conditions in detail, as it is relevant for our numerics. 

Free unit cells are the unit cells that can be shifted between any pair of domain walls via domain wall hops on the lattice. Leaving the calculation of the number of these free cells aside for the moment, the Krylov sector dimension is the number of arrangements of $n_{\mathrm{f}}$ free unit cells between $n_{\mathrm{DW}}$ domain walls, which is given by
\begin{equation}\label{eq:krylov_space_dim_obc}
    D_{K, \mathrm{OBC}} = 
    \binom{n_{\mathrm{f}} + n_{\mathrm{DW}}}{n_{\mathrm{f}}}.
\end{equation}
This equation can be understood by considering a chain of $n_{\mathrm{f}} + n_{\mathrm{DW}}$ sites. The number of ways of picking $n_{\mathrm{f}}$ sites to put free cells into is given by \autoref{eq:krylov_space_dim_obc}. The remaining sites contain the domain walls, the arrangement of which is fully determined by the reference string. 

With periodic boundary conditions, the counting is altered because reference strings related by cyclic permutations of the domain walls are identified with one other. Therefore, the number of unique reference strings that define Krylov sectors is reduced compared to open boundary conditions, and the dimension of each sector is larger. 

Fix one of the domain walls to a fixed bond in the spin chain. The number of configurations in the Krylov sector with this fixed domain wall is the number of ways of arranging $n_{\mathrm{f}} \geq 1$ unit cells between the remaining $n_{\mathrm{DW}}-1 \geq 0$ domain walls. (When \(n_{\mathrm{f}} = 0\), there are no possible hops of domain walls, and \(D_K = 1\). Similarly, \(n_{\mathrm{DW}} = 0\) corresponds to one of the N\'eel states, for which \(D_K = 1\).) Naively, the first domain wall may be in any one of $L/2$ positions ($L/2$ instead of $L$ because its type, $\mathrm{i}$ or $\mathrm{o}$ is fixed). However, as domain walls of fixed type are indistinguishable particles, we need to divide by $n_{\mathrm{stab}}$, where $n_{\mathrm{stab}}$ is the number of cyclic permutations of the reference string that leave it invariant. All together, 
\begin{equation}\label{eq:krylov_space_dim}
    D_{K} = \binom{n_{\mathrm{f}} + n_{\mathrm{DW}}-1}{n_{\mathrm{f}}} \times \frac{L}{2 n_{\mathrm{stab}}}. 
\end{equation}
An example with $n_{\mathrm{stab}}$ > 1 is the reference string of $\psi_{\mathrm{DW2}}$. There are two cyclic permutations that leave the string $`+_{\mathrm{o}}, -_{\mathrm{o}}, +_{\mathrm{o}}, -_{\mathrm{o}}$' invariant: the identity and two cyclic shifts.

We now derive a formula for the number of free unit cells $n_{\mathrm{f}}$ for periodic boundary conditions.
All unit cells of the spin lattice can be divided into free and bound cells (\autoref{fig:dw_sites}). The bound cells are fixed to the domain walls. We first count the number of bound cells $n_{\mathrm{b}}$ and then use the relationship $L/2 = n_{\mathrm{b}} + n_{\mathrm{f}}$ to count the free cells. 
Bound cells are either occupied by the $\pm_{\mathrm{i}}$ domain walls (\figref{fig:dw_sites}{a}), or are single unit cells between adjacent pairs of $\mathrm{o}$ domain walls. For example, there are four cells between adjacent pairs of $\mathrm{o}$ domain walls in \figref{fig:dw_sites}{b}, including between the last and first domain wall. We define the rightmost cell between an adjacent pair of $\mathrm{o}$ domain walls as the bound unit cell, which we imagine is attached to the domain wall to its right. This convention is chosen as the $\pm 1$ sign of the domain wall associated with the bound cell fixes its imbalance (\figref{fig:dw_sites}{a}). The total number of bound cells in a domain wall string is 
\begin{equation}\label{eq:bound_cells}
    n_{\mathrm{b}} = n_{\mathrm{DW,i}} + p_{\mathrm{o}},
\end{equation}
where $n_{\mathrm{DW,i}}$ counts the number of $\mathrm{i}$ domain walls and $p_{\mathrm{o}}$ the number of adjacent pairs of $\mathrm{o}$ domain walls. Thus, the number of free unit cells $n_{\mathrm{f}}$ is
\begin{equation}\label{eq:free_cells}
    n_{\mathrm{f}} = L/2 - n_{\mathrm{DW,i}} - p_{\mathrm{o}}.
\end{equation}
Therefore, given a reference string and an even number of lattice sites $L/2$, \autoref{eq:krylov_space_dim} can be used to read off the dimension of the Krylov sector as shown in the caption of \autoref{fig:dw_sites}.

\subsection{Krylov Sector Imbalance}\label{sec:imbalance}

In this section, we derive the average imbalance formula for all Krylov sectors with $S^{z}_{\mathrm{tot}}=0$ and \(n_{\mathrm{f}} \geq 1\). 

Since free unit cells contribute $\pm 1$ to imbalance, we enumerate all configurations with a particular number of $-1$ contributing free cells $n_-$, and subsequently sum over all possible values of $n_-$. 
Above, we established that the states in the sector are obtained by distributing free cells between the domain walls. The imbalance contribution of free cells is \(+1\) $(-1)$ if they are to the left of a domain wall of type $+(-)$. As free cells do not contain domain walls, and so have nonzero imbalance, we have the relation $n_{\mathrm{f}}=n_+ + n_-$. 
While the bound cells with $\mathrm{i}$ domain walls in them have $0$ imbalance, the bound cells between the $p_{\mathrm{o}}$ adjacent pairs of $\mathrm{o}$ domain walls have a fixed imbalance determined from the reference string. As a reminder, the imbalances of these bound cells are given by the sign of the $\mathrm{o}$ domain wall to their right-hand side. We denote the imbalance contribution from such bound cells as $S(j)=\pm 1$, where $j\in\{1, ... , p_\mathrm{o}\}$ indexes bound cells with $\pm_{\mathrm{o}}$ domain wall to their right hand side. 

The imbalance of the state with uniform occupation of all $z$-basis states in a Krylov sector is
\begin{equation}\label{eq:I_Kss}
\begin{split}
    I_{\mathrm{ss}} = &\frac{1}{D_K} \sum_{n_{-}=0}^{n_\mathrm{f}} \left[ n_+-n_{-}+\sum_{j=1}^{p_{\mathrm{o}}}S(j) \right] \times \\
    & {{\frac{n_{\mathrm{DW}}}{2} - 1+n_+}\choose{n_+}} {{\frac{n_{\mathrm{DW}}}{2} - 1+n_{-}}\choose{n_{-}}}.
\end{split}
\end{equation}
The term in the square bracket is the imbalance of a given configuration $n_{+}-n_{-} + \sum_j S(j)$, where the offset $\sum_j S(j)$ results from the imbalance of the bound unit cells as described above. For the state $\psi_{\mathrm{DW1}}$ the offset is $-1$ since there is one adjacent pair of $\mathrm{o}$ walls with $-_{\mathrm{o}}$ on the right \figref{fig:dw_sites}{a}. However, $\psi_{\mathrm{DW2}}$ state has offset zero, as four paired $\mathrm{o}$ domain walls cancel each other's imbalance contributions. The combinatorial factors in~\autoref{eq:I_Kss} account for the number of ways to distribute the $n_+$ and $n_-$ free unit cells between the $\frac{n_{\mathrm{DW}}}{2}$ number of $+1$ and $-1$ imbalance regions, respectively.

The average imbalance of a Krylov sector is non-zero if the offset in the bracket on the first line of \autoref{eq:I_Kss} is non-zero. Since the combinatorial factors are symmetric in the exchange of $n_+$ and $n_-$, $I_{\mathrm{ss}}=0$ if the imbalance contributions from the paired o domain walls cancel. If $p_{\mathrm{o}}$ is odd, the average imbalance is non-zero. The average imbalance can also be non-zero for even $p_{\mathrm{o}}$ if the offset contributions of the paired domain walls do not cancel. For example, the string $`+_{\mathrm{o}}, -_{\mathrm{o}},+_{\mathrm{i}}, -_{\mathrm{i}},+_{\mathrm{o}}, -_{\mathrm{o}},+_{\mathrm{i}}, -_{\mathrm{i}}$' has $p_{\mathrm{o}}=2$, both with negative imbalance bound cells. The state $\psi_{\mathrm{DW1}}$ has an imbalance bias because $p_{\mathrm{o}}=1$. Using \autoref{eq:I_Kss}, we find the average imbalance in the $\psi_{\mathrm{DW1}}$ Krylov sector is $I_{\mathrm{ss}} = -1/8$. However, the state $\psi_{\mathrm{DW2}}$ has $p_{\mathrm{o}}=4$ with cancelling offset contributions, and thus zero average imbalance for its Krylov sector.

\section{Discussion}

This article demonstrates how applying controlled dephasing to Hilbert space fragmented systems enables simple probes of Krylov sector properties. Specifically, we study the strongly interacting XXZ chain subject to dephasing noise and find clear signatures of Krylov sector properties in the quench dynamics of the sublattice spin imbalance when the system is approximately fragmented. 

Namely, when initialized in a state with positive imbalance that belongs to a sector with an average negative imbalance, there is a striking overshoot of the imbalance as a function of time past the maximally mixed expectation of zero.
At large interaction and dephasing strengths, $\Delta \gg \gamma \gg J$, the average imbalance of Krylov sectors can be read off from this imbalance overshoot. 

Krylov sector properties are less pronounced in the coherent dynamics (for $\gamma=0$ and $\Delta \gg J$) and, therefore, more challenging to extract. On $\mathcal{O}(J^{-1})$ timescales, the system exhibits imbalance oscillations around the average imbalance value of the associated sector. The difficulty in predicting the size and scale of the oscillations obscures the measurement of the Krylov sector average.
With dephasing, the overshoot occurs on timescales $T_{\mathrm{hop}} = \order{\gamma J^{-2}}$, which can be tuned by controlling $\gamma/J$. Thus, introducing dephasing provides additional control over the dynamics, enabling simpler extraction of Krylov sector properties.

Our proposal is realistic in ultracold atomic setups, such as the spinful Fermi-Hubbard model~\citep{scherg2021observing_Exp_1d_hubbard, kohlert2023exploring_1d_hubbard}. The interaction strengths necessary to extract an imbalance overshoot that reflects the average imbalance of the sector are achievable in these systems~\citep{RevModPhys.80.885, scherg2021observing_Exp_1d_hubbard, kohlert2023exploring_1d_hubbard}. Our results also do not rely on translational or global-spin flip invariance. Therefore, experimental imperfections in realizing the XXZ model in the form of weak disorder in the $S^z$ field would leave our conclusions intact. On the other hand, particle loss and other number-conservation-breaking noise spoil fragmentation in the XXZ model, and its presence imposes a timescale beyond which experimental measurements do not reflect the fragmented model.

A potential extension of this work is to explore other systems that are fragmented in the computational basis, since controlled dephasing in this basis can be naturally applied in experimental platforms.
We expect most observables can act as witnesses of fragmentation, as imbalance does for the strongly interacting XXZ model. 
Furthermore, engineered decoherence may allow similar methods to be applied to more general models that are not fragmented in the computational basis~\citep{moudgalya2022hilbert_HSF, li2023hilbert_HSF}.

Another direction is to understand imbalance dynamics starting from the Neel state in the presence of controlled dephasing. In the fragmented limit ($\Delta \to \infty$), the Neel state is frozen with or without dephasing. Numerical simulations at large finite $\Delta$ show that the imbalance exhibits an overshoot, much like $\psi_{\mathrm{DW1}}$, albeit with a much smaller size of the overshoot. Explaining the size and sign of this overshoot through the physics of Krylov sectors requires a treatment of the nucleation events. It would also be interesting to revisit previous work~\cite{everest2017role} that predicts compressed exponential decay for the imbalance with HSF in mind.

\acknowledgements
The authors thank A. Khoudorozhkov and M. Aidelsburger for helpful discussions. This work was supported by: NSF Grant No. DMR-1752759 (D.V. and A.C.); AFOSR Grant no. FA9550-24-1-0121 (A.C. and R.S.), a Stanford Q-FARM Bloch Fellowship (DML) and a Packard Fellowship in Science and Engineering (DML, PI: Vedika Khemani). A.C. acknowledges the hospitality of the Max Planck Institute for Complex Systems. Numerical work was performed on the BU Shared Computing Cluster. 

\bibliography{ref.bib}

\appendix

\section{Non-interacting limit}\label{sec:non_int_limit}

This appendix describes the non-interacting limit of the model in \autoref{eq:Lindblad_equation}. At $\gamma=0$, the imbalance as a function of time for either initial state $\psi_{\mathrm{DW1}}$ and $\psi_{\mathrm{DW2}}$ is visually indistinguishable, and displays power law decaying oscillations at long times. For increasing dephasing rate $\gamma > 0$, the imbalance oscillations decay exponentially in time, and eventually become overdamped at some numerically observed dephasing rate $\gamma = \mathcal{O}(J)$.

\begin{figure}
    \centering
    \includegraphics[width=\linewidth]{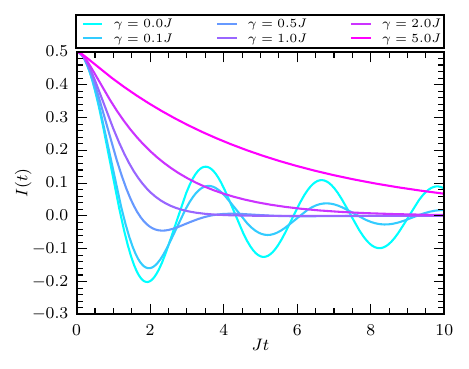}
    \caption{Imbalance curves in the dephased XX model as a function of $\gamma/J$ in the $\psi_{\mathrm{DW2}}$ initial state. We construct a $L=160$ size chain by repeating the pattern of the $\psi_{\mathrm{DW2}}$ initial state 10 times. In the non-interacting limit, the dynamics of the $\psi_{\mathrm{DW1}}$ and $\psi_{\mathrm{DW2}}$ states are identical. Without dephasing, the imbalance curve is the exactly known zeroth order Bessel function of the first kind. Weak dephasing $\gamma < J$ results in underdamped imbalance oscillations, and an exponentially decaying envelope is scaling with $\gamma$ (\autoref{fig:damp}). At large dephasing $\gamma > J$, the imbalance curves decay exponentially with a rate scaling as $J^2/\gamma$. Curves are extracted using the method in~\citep{vznidarivc2010exact, vznidarivc2013transport} that exploits the decoupling of the fermionic 2-point correlation functions from n-point correlations in the Jordan-Winger transformed version of the dephased system~\autoref{eq:Lindblad_equation} with $\Delta=0$ and $J=1$.} 
    \label{fig:non-int}
\end{figure}

The closed system imbalance dynamics, at $\gamma=0$, can be computed exactly by mapping to free fermions using a Jordan-Wigner transformation. The imbalance as a function of time is given by~\citep{LIEB1961407}
\begin{equation}\label{eq:Bessel} 
    I(t) = \frac{2}{L}\sum_{m_{\uparrow}} e^{i \pi m_{\uparrow}} \mathcal{J}_0(2Jt),
\end{equation}
where $\mathcal{J}_0(2Jt)$ is the zeroth order Bessel function of the first kind and the sum is over the positions $m_{\uparrow}$ of all spins pointing upwards in the initial state (or equivalently, particles in the fermion language). After time $t=\mathcal{O}\left( J^{-1} \right)$ the~\autoref{eq:Bessel} in the N\'eel initial state becomes 
\begin{equation}
    I(t) \sim \frac{1}{\sqrt{4\pi t}} \cos(2Jt).
\end{equation}
The intuitive picture for imbalance oscillations through the ballistic spread of wavepackets on the chain has been explained in \autoref{sec:results_unitary}.

\begin{figure}[b]
    \centering
    \includegraphics[width=\linewidth]{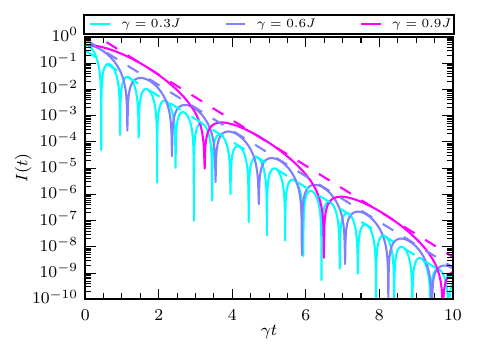}
    \caption{Evidence for the underdamped imbalance oscillations in the dephased XX chain. We show the envelope decay of the imbalance oscillations in the underdamped regime $\gamma < J$ when the system is initialized in the $\psi_{\mathrm{DW2}}$ state. All the dashed lines going through the peaks of the imbalance curves have the same slope, indicating that the decay rate of the oscillation envelopes is proportional to $\gamma$. \emph{Parameters}: $\Delta=0$, $J=1$ and $L=160$.}
    \label{fig:damp}
\end{figure}

A non-zero dephasing rate can be viewed as randomly performing measurements on the system in the \(S^z\) basis. This measurement annihilates the coherent superposition of wavepackets. When $\gamma \ll J$, the measurements are performed at a lower frequency than the imbalance oscillations, and the oscillatory behavior persists. Indeed, in \autoref{fig:non-int} we numerically observe that the oscillations persist, but become damped, with an exponentially decaying envelope whose decay rate scales with $\gamma$ (\autoref{fig:damp}). For dephasing rates $\gamma \gtrsim J$, the oscillations can no longer be observed---we enter the overdamped regime, with exponential imbalance decay at longer times. Here, the decay rate scales as $J^2/\gamma$---decreasing with the dephasing rate. This is an example of the quantum Zeno effect~\citep{misra1977zeno}, where rapidly measuring the system in the $z$-basis slows down the decay of observables composed of $S^{z}$ operators. The exponential imbalance decay can be seen in the effective model \autoref{eq:effective_Hamiltonian}, where the only timescale in the model is $J_{\mathrm{hop}}=J_{\mathrm{nuc}}=J^2/\gamma$. This implies that on timescales $t \gg 1/\gamma$ the imbalance decays as $e^{-J^2 t/\gamma}$.

Note that it is precisely the overdamped regime that we are interested in when discussing the imbalance overshoot in the strongly interacting system in \autoref{sec:inf_int_dephas}. Dephasing causes the imbalance oscillations in the strongly interacting regime to decay with increasing $\gamma$. This can be seen by increasing $\gamma/J$ for fixed $\Delta/J$ in \figref{fig:phase_diag}{b} and \figref{fig:phase_diag}{c}. At some $\gamma = \mathcal{O}(J)$, the system becomes overdamped and oscillations do not occur.

\end{document}